\documentclass[12pt]{iopart}

\usepackage{cite}
\usepackage[utf8]{inputenc}
\usepackage{graphicx}
\usepackage{graphicx,epstopdf}
\usepackage{dcolumn}
\usepackage{amsmath}
\usepackage{lineno,hyperref}
\usepackage{amsfonts}
\usepackage{amsthm}
\usepackage{bbm}
\usepackage{amssymb}
\usepackage{mathrsfs}
\PassOptionsToPackage{caption=false}{subfig}
\usepackage{subfig}
\usepackage{color}
\usepackage[normalem]{ulem}

\usepackage{blindtext}
\usepackage{lipsum}
\usepackage{bbm}

\usepackage{etoolbox}

\makeatletter
\newcommand{\mainmatter}{%
	\setcounter{footnote}{0}%
	\patchcmd{\@makefntext}{\fnsymbol}{\arabic}{}{}%
	\patchcmd{\@thefnmark}{\fnsymbol}{\arabic}{}{}%
	\def\@makefnmark{\textsuperscript{\arabic{footnote}}}%
}
\makeatother

\begin{document}
\title[]{Quantum fluctuation theorem for initial near-equilibrium system}

\author{Bao-Ming Xu}

\address{Shandong Key Laboratory of Biophysics, Institute of Biophysics, Dezhou University, Dezhou 253023, China}

\ead{xbmv@bit.edu.cn}

\vspace{10pt}
\begin{indented}
\item[]
\end{indented}

\begin{abstract}
Quantum fluctuation theorem commonly requires the system initially prepared in an equilibrium state. Whether there exists universal exact quantum fluctuation theorem for initial states beyond equilibrium needs further discussions. In the present paper, we initialize the system in a near-equilibrium state, and derive the corresponding modified Jarzynski equality by using perturbation theory. The correction is nontrivial since it directly leads to the principle of maximum work or the second law of thermodynamics for near-equilibrium system, and also offers a much tighter bound of work. Two prototypical near-equilibrium systems driven by a temperature gradient and an external field, are taken into account, to confirm the validity and the generality of our theoretical results. Finally, a fundamental connection between quantum critical phenomenon and near-equilibrium state at really high temperature is revealed.
\end{abstract}

\vspace{2pc}

\maketitle

\section{Introduction}\label{sec1}
When the size of a physical system is scaled down to the micro-/nano-scopic domain, fluctuations of relevant quantities start playing a pivotal role in establishing the energetics of the system; therefore, the laws of thermodynamics have to be given by taking into account the effects of these fluctuations \cite{Seifert2012,Esposito2009,Campisi2011,Marconia2008,Kubo1966,Callen1951}. This line of research dates back to Einstein and Smoluchowski, who derived the connection between fluctuation and dissipation effects for Brownian particles \cite{Marconia2008}. Now, it is well known that near-equilibrium, linear response theory provides a general proof of a universal relation known as the fluctuation-dissipation theorem (FDT), which states that the response of a given system when subject to an external perturbation is expressed in terms of the fluctuation properties of the system in thermal equilibrium \cite{Seifert2012,Marconia2008,Kubo1966,Callen1951}. It offers a powerful tool to analyze general transport properties in numerous areas, from hydrodynamics to many-body and condensed-matter physics. Over the past few decades, FDT has been successfully generalized to nonequilibrium steady state (NESS) in both classical \cite{Speck2006,Seifert2010,Chetrite2008,Baiesi2009,Qian2006,Prost2009,Baiesi2013,Ciliberto2010,Gingrich2016,Znidaric2019} and quantum \cite{Zhang2016,Wang2021,Konopik2019,Mehboudi2018} systems. Beyond this linear response regime, for a long time, no universal exact results were available due to the detailed balance breaking.

The breakthrough came with the discovery of the exact fluctuation theorems (FTs) \cite{Seifert2012,Esposito2009,Campisi2011}, which hold for the system arbitrarily far from equilibrium and reduce to the known FDTs for the system near-equilibrium. One of the most important FTs is Jarzynski equality \cite{Jarzynski1997,Tasaki2000,Kurchan2000,Mukamel2003}:
\begin{equation}\label{JE}
  \langle e^{-\beta W_e}\rangle=e^{-\beta \Delta F_e},
\end{equation}
where $\langle\circ\rangle$ denotes the average over an ensemble of measurements of fluctuating work $W_e$ in a nonequilibrium process:
\begin{equation}\label{}
  \lambda_0\longrightarrow\lambda_\tau~~\textmd{or}~~H(\lambda_0)\longrightarrow H(\lambda_\tau),
\end{equation}
i.e., the work parameter is changing from its initial value $\lambda_0$ to the final value $\lambda_\tau$ or the system Hamiltonian is changing from $H(\lambda_0)$ to $H(\lambda_\tau)$. In Eq. (\ref{JE}) 
\begin{equation}\label{}
\Delta F_e=-T\ln\frac{Z(\lambda_\tau)}{Z(\lambda_0)}
\end{equation}
is Helmholtz free energy difference between initial equilibrium state
\begin{equation}\label{}
  \rho_{e}(\lambda_0)\equiv\frac{e^{-\beta H(\lambda_0)}}{Z(\lambda_0)}.
\end{equation}
and final equilibrium state $\rho_{e}(\lambda_\tau)=e^{-\beta H(\lambda_\tau)}/Z(\lambda_\tau)$, where $Z(\lambda_t)=\mathrm{Tr}\{\exp[-\beta H(\lambda_t)]\}$ refers to the partition function with respect to Hamiltonian $H(\lambda_t)$ and the external time dependent work parameters held fixed at $\lambda_t$, $\beta\equiv1/T$ is the inverse temperature (we set Boltzmann constant $k_B=1$ throughout this paper). It should be noted that, like the condition of FDT, the validity of Jarzynski equality also requires the system initially prepared in the equilibrium state $\rho_{e}(\lambda_0)$. In other words, Jarzynski equality only holds for a process starting from an equilibrium state. We have introduced subscript $e$ to stand for equilibrium, and it will remain for the rest of discussions. Whether there exist universal exact FTs for a process starting from a state beyond equilibrium? In classical system, some FTs can be generalized to NESS \cite{Evans1993,Gallavotti1995,Kurchan1998,Lebowitz1999,Seifert2005,Searles2007,Hatano2001,Solano2010,Mounier2012,Chetrite2008,Liu2010}, even to an arbitrary state \cite{Gong2015,Crooks2000}, while in the quantum regime, the situation is more subtle and challenging.

With the aid of quantum information theory, a resource theory framework of thermodynamics has been established \cite{Goold2016}. In such resource theoretic framework \cite{Brandao2013,Gour2015,Streltsov2017,Chitambar2019,Lostaglio2019}, the system dynamics is simulated by thermal operations \cite{Kraus1983,Janzing2000,Horodecki2013}, where the external work protocol is performed by including a switch or some other systems, and work is considered as the change in the energy of some work systems or weights. With these thermodynamic concepts at hand, some FTs are generalized to the state beyond equilibrium \cite{Aberg2018,Holmes2019,Kwon2019}, where quantum coherence is shown to catalyze the system state transformations under some thermodynamical constrains \cite{Alhambra2016,Alhambra2016_1,Aberg2014,Mingo2019,Morris2018}. It should be noted that those additional interactions will inevitably affect the system, which can be understood as some effective measurements, and thus the resource theory results can not be applied to the non-equilibrium process described only from the system itself, such as the unitary process Eq. (\ref{UU}) considered in this paper. Can we generalize FTs only from the system itself? To the best of our knowledge, it is still unknown, and the first main goal of this paper is to answer this question.

\begin{figure}
\begin{center}
\includegraphics[width=8cm]{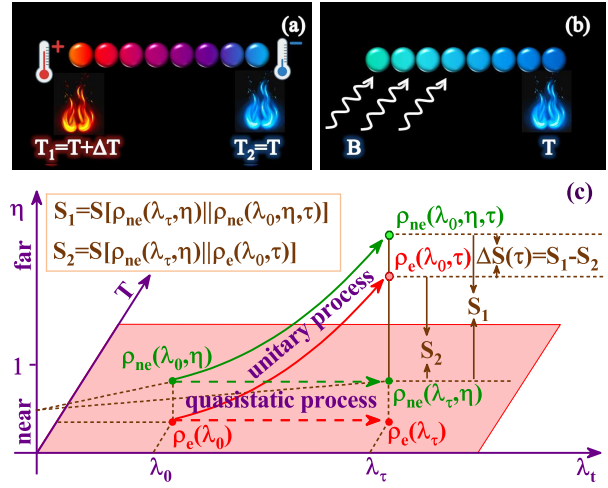}
\caption{(Color online) Sketch of a near-equilibrium system (a chain as an example): (a) The system is simultaneously coupled to two heat reservoirs with temperatures $T+\Delta T$ and $T$, respectively; (b) The system coupled to a heat reservoir with temperature $T$ is driven away from equilibrium state by a weak field $B$. (c) General schematics for the contrast of equilibrium and near-equilibrium states, and the contrast of quasistatic and unitary processes. A system with Hamiltonian $H(\lambda_t)$ weakly coupled to an environment with temperature $T$ will eventually stabilize at equilibrium state $\rho_{e}(\lambda_t)= e^{-\beta H(\lambda_t)}/Z(\lambda_t)$, thus the equilibrium state can be completely determined by the given external parameter $\lambda_t$ and temperature $T$. All the equilibrium states compose the equilibrium manifold, i.e., the plane (pink) consisting of $\lambda_t$ and $T$ axes. In addition to $\lambda_t$ and $T$, the determination of near-equilibrium states $\rho_{ne}(\lambda_t,\eta)$ requires another parameter $\eta$, which is relatively small. In equilibrium thermodynamics, a process that the system always remains in the equilibrium state (red dashed arrow) is called quasistatic process. Here, we assume a process that the system always remains in the near-equilibrium state (green dashed arrow), e.g., external parameter $\lambda_t$ is infinite-slowly changed from $\lambda_0$ to $\lambda_\tau$ with a fixed small $\eta$, is also called quasistatic process. A unitary process $U(\tau)$ changing the external parameter from $\lambda_0$ to $\lambda_\tau$ (red and green solid arrows) can transfer the system from near-equilibrium state $\rho_{ne}(\lambda_0,\eta)$ [or equilibrium state $\rho_e(\lambda_0)$] to $\rho_{ne}(\lambda_0,\eta,\tau)=U(\tau)\rho_{ne}(\lambda_0,\eta)U^\dag(\tau)$ [or $\rho_e(\lambda_0,\tau)=U(\tau)\rho_e(\lambda_0)U^\dag(\tau)$] which may be far away from equilibrium. The distances between the far-from-equilibrium and the near-equilibrium states are described by relative entropy $S_1$ and $S_2$. The difference of them, i.e., $\Delta S(\tau)=S_1-S_2$, is used to modify Jarzynski equality and the principle of maximum work or the second law of thermodynamics so that they apply to the initial near-equilibrium systems. It should be noted that $\Delta S(\tau)\neq S[\rho_{ne}(\lambda_0,\eta,\tau)||\rho_e(\lambda_0,\tau)]$. }
\label{figure1}
\end{center}
\end{figure}

Using Jensen's inequality $\langle e^x\rangle\geq e^{\langle x\rangle}$, Jarzynski equality yields the principle of maximum work:
\begin{equation}\label{e2thlaw}
\langle W_e\rangle\geq\Delta F_e.
\end{equation}
In this sense, Jarzynski equality can be also regarded as one of the fundamental generalizations of the second law of thermodynamics. The equal sign holds for the infinitely slow quasistatic process starting from one equilibrium state to another, i.e, equilibrium free energy difference $\Delta F_e$ is the average work in the infinitely slow quasistatic process. Any finite-time process will do more work, but the work beyond the equilibrium free energy difference, namely, the irreversible work, will ultimately dissipate to the environment. In other words, the infinitely slow quasistatic process is the most efficient because no extra energy is wasted. Here we posit a fundamental question: For all the processes from one NESS to another, is the quasistatic reversible one still the most efficient? Or how to describe the second law of thermodynamics beyond equilibrium? To discuss the second law of thermodynamics beyond equilibrium is the second main goal of this paper.

In this paper we consider a quantum system is driven away from equilibrium by a priori small perturbation, e.g., small temperature gradient [see Fig. \ref{figure1}(a)] or weak external field [see Fig. \ref{figure1}(b)]. After that, the system will stabilize at a near-equilibrium state. We then consider a general post work protocol performed on the system and investigate quantum work fluctuations.

\section{Near-equilibrium state}\label{sec2}
Let us begin with the elaboration of the near-equilibrium state. The near-equilibrium state is a steady state obtained by: for example, (1) The system is simultaneously coupled with two heat reservoirs at temperatures $T+\Delta T$ and $T$, respectively [see Fig. \ref{figure1}(a)], where temperature difference $\Delta T$ is sufficiently small; or (2) the system coupled to a heat reservoir at temperatures $T$ is driven away from equilibrium by a weak field $B$ [see Fig. \ref{figure1}(b)]. For the small temperature difference or the weak field driving, we assume the system stabilizes at the near-equilibrium state.

Given an initial Hamiltonian
\begin{equation}\label{}
H(\lambda_0)=\sum_mE_m(\lambda_0)|E_m(\lambda_0)\rangle\langle E_m(\lambda_0)|
\end{equation}
and a temperature $T$, the near-equilibrium state has a spectral decomposition:
\begin{equation}\label{rhone1}
\rho_{ne}(\lambda_0,\eta)=\sum_m\rho_{mm}(\lambda_0,\eta)
|\psi_m(\lambda_0,\eta)\rangle\langle\psi_m(\lambda_0,\eta)|.
\end{equation}
Here, and in all the following discussions, subscript $ne$ stands for near-equilibrium. $\eta$ is a quantity that quantifies the intensity of perturbation. For case (1), $\eta=\Delta T/T<1$ is the relative temperature difference; for case (2), $\eta=B/\mathcal{E}(\lambda_0)<1$ is the relative driving strength with $\mathcal{E}(\lambda_0)$ being the overall energy scale of the system. If $\eta=0$ (i.e., $\Delta T=0$ or $B=0$), the system is stable at equilibrium state, i.e., $\rho_{ne}(\lambda_0,0)=\rho_{e}(\lambda_0)$. In this sense, $\eta$ also quantifies the differences between the near-equilibrium state and the equilibrium state. Using perturbation theory, $\rho_{mm}(\lambda_0,\eta)$ and $|\psi_m(\lambda_0,\eta)\rangle$ can be expressed as
\begin{equation}\label{rhommeta}
  \rho_{mm}(\lambda_0,\eta)\approx\frac{e^{-\beta E_m(\lambda_0)}}{Z(\lambda_0)}+\rho'_{mm}(\lambda_0,0)\eta
\end{equation}
and
\begin{equation}\label{phimeta}
  |\psi_m(\lambda_0,\eta)\rangle\approx|E_m(\lambda_0)\rangle+\eta|\psi'_m(\lambda_0,0)\rangle,
\end{equation}
where $\rho'_{mm}(\lambda_0,0)\equiv\frac{\partial\rho_{mm}(\lambda_0,\eta)}{\partial\eta}\big\vert_{\eta=0}$ and $|\psi'_m(\lambda_0,0)\rangle\equiv\frac{\partial|\psi_m(\lambda_0,\eta)\rangle}{\partial\eta}\big\vert_{\eta=0}$. In this paper, we only consider the first order of $\eta$. Substituting Eq. (\ref{rhommeta}) and Eq. (\ref{phimeta}) into Eq. (\ref{rhone1}), the near-equilibrium state can be expressed as
\begin{equation}\label{}
  \rho_{ne}(\lambda_0,\eta)=\rho_e(\lambda_0)+\rho'_{ne}(\lambda_0,0)\eta
\end{equation}
with
\begin{equation}
\rho'_{ne}(\lambda_0,0)=\sum_m\rho'_{mm}(\lambda_0,0)|E_m(\lambda_0)\rangle\langle E_m(\lambda_0)|
 +\frac{e^{-\beta E_m(\lambda_0)}}{Z(\lambda_0)}\Bigl[|E_m(\lambda_0)\rangle\langle\psi'_m(\lambda_0,0)|+h.c.\Bigr],
\end{equation}
where $h.c.$ means the Hermitian conjugate expression. Because $\mathrm{Tr}[\rho_{ne}(\lambda_0,\eta)]=\mathrm{Tr}[\rho_e(\lambda_0)]=1$, $\mathrm{Tr}[\rho'_{ne}(\lambda_0,0)]=0$.

According to the normalization of $\langle\psi_m(\lambda_0,\eta)|\psi_m(\lambda_0,\eta)\rangle=\langle E_m(\lambda_0)|E_m(\lambda_0)\rangle=1$, $|\psi'_m(\lambda_0,0)\rangle$ and $|E_m(\lambda_0)\rangle$ satisfy
\begin{equation}\label{}
\langle E_m(\lambda_0)|\psi'_m(\lambda_0,0)\rangle
+\langle \psi'_m(\lambda_0,0)|E_m(\lambda_0)\rangle=0.
\end{equation}
Using this property, one can found that
\begin{equation}\label{pm}
  \langle E_m(\lambda_0)|\rho_{ne}(\lambda_0,\eta)|E_m(\lambda_0)\rangle
  \approx\frac{e^{-\beta E_m(\lambda_0)}}{Z(\lambda_0)}+\eta\rho'_{mm}(\lambda_0,0)
  \approx\rho_{mm}(\lambda_0,\eta),
\end{equation}
i.e., the near-equilibrium energy populations are approximately equal to the eigenvalues of near-equilibrium state. This idea of using perturbation theory to approximate the thermodynamics of an non-equilibrium state to that of an equilibrium state has also been explored in the framework of response theory \cite{Konopik2019,Mehboudi2018,Hsiang2020}.

In addition to the requirement of small perturbation, i.e., $\eta<1$, near-equilibrium state also requires energy populations do not change significantly from the equilibrium state, i.e.,
\begin{equation}\label{}
\Bigg\vert\frac{\rho_{mm}(\lambda_0,\eta)-\frac{e^{-\beta E_m(\lambda_0)}}{Z(\lambda_0)}}{\frac{e^{-\beta E_m(\lambda_0)}}{Z(\lambda_0)}}\Bigg\vert\approx
\bigg\vert\frac{\eta\rho'_{mm}(\lambda_0,0)}{\frac{e^{-\beta E_m(\lambda_0)}}{Z(\lambda_0)}}\bigg\vert<1
\end{equation}
holds for all $m$. This second requirement is not redundance. Given a system energy level $E_m(\lambda_0)$, its equilibrium population $\rho_{mm}(\lambda_0,0)=\frac{e^{-\beta E_m(\lambda_0)}}{Z(\lambda_0)}$ is completely determined by temperature or thermal fluctuations, thus $\Big\vert\eta\rho'_{mm}(\lambda_0,0)/\frac{e^{-\beta E_m(\lambda_0)}}{Z(\lambda_0)}\Big\vert<1$ essentially implies perturbation should not be stronger than thermal fluctuations. In case (1), $\eta=\Delta T/T<1$ has told us that the perturbation is always weaker than thermal fluctuation; but in case (2), the competition between weak field and thermal fluctuations has not been discussed. Given a perturbation $\Delta T$ or $B$, near-equilibrium state $\rho_{ne}(\lambda_0,\eta)$ and equilibrium state $\rho_e(\lambda_0)$ will both approach to the maximally mixed state if the system goes to sufficiently high temperature, i.e., $\lim_{T\rightarrow\infty}\rho_{ne}(\lambda_0,\eta)=\lim_{T\rightarrow\infty}\rho_{e}(\lambda_0)=\mathbb{I}/N$, where $\mathbb{I}$ is the identity matrix and $N$ is the total degree of freedom of the system. In this case, $\rho'_{mm}(\lambda_0,0)\rightarrow0$, and thus $\Big\vert\eta\rho'_{mm}(\lambda_0,0)/\frac{e^{-\beta E_m(\lambda_0)}}{Z(\lambda_0)}\Big\vert\rightarrow0$, satisfying the second requirement of near-equilibrium state. If the system goes to low temperature, the equilibrium populations of the high energy levels will approach to zero. In this case, the second requirement $\Big\vert\eta\rho'_{mm}(\lambda_0,0)/\frac{e^{-\beta E_m(\lambda_0)}}{Z(\lambda_0)}\Big\vert<1$ can be violated, and the system is far away from equilibrium. This can be understood as follows, at low temperature, thermal fluctuations are suppressed, and the system is so sensitive to external perturbation that it will be far away but not near from equilibrium.

The second requirement of near-equilibrium state can also be described from other perspective. Because $\Big\vert\eta\rho'_{mm}(\lambda_0,0)/\frac{e^{-\beta E_m(\lambda_0)}}{Z(\lambda_0)}\Big\vert<1$,
\begin{equation}\label{}
  \Bigg\vert\ln\biggl(1+\frac{\rho'_{mm}(\lambda_0,0)\eta}{\frac{e^{-\beta E_m(\lambda_0)}}{Z(\lambda_0)}}\biggr)\Bigg\vert
  \approx\Bigg\vert\ln\frac{\rho_{mm}(\lambda_0,\eta)}{\frac{e^{-\beta E_m(\lambda_0)}}{Z(\lambda_0)}}\Bigg\vert<1
\end{equation}
holds for all $m$. In this regards, one can define an arbitrary probability distribution $\{p_m\}$ that makes
\begin{equation}\label{}
  \Bigg\vert\sum_mp_m\biggl[\ln\rho_{mm}(\lambda_0,\eta)-\ln\frac{e^{-\beta E_m(\lambda_0)}}{Z(\lambda_0)}\biggr]\Bigg\vert<1
\end{equation}
hold. Translating into the language of quantum mechanics, one can define an arbitrary third state $\varrho$, namely, the reference state, to make
\begin{equation}
\Big\vert\mathrm{Tr}[\varrho\ln\rho_{ne}(\lambda_0,\eta)]-\mathrm{Tr}[\varrho\ln\rho_e(\lambda_0)]\Big\vert
=\Big\vert S[\varrho||\rho_{ne}(\lambda_0,\eta)]-S[\varrho||\rho_e(\lambda_0)]\Big\vert
=\big\vert\Delta S\big\vert <1
\end{equation}
hold, where $S[\varrho||\rho]=\mathrm{Tr}[\varrho\ln\varrho]-\mathrm{Tr}[\varrho\ln\rho]$ is the relative entropy between $\varrho$ and $\rho$. $|\Delta S|$ can be understood as the indirect distance between $\rho_{ne}(\lambda_0,\eta)$ and $\rho_e(\lambda_0)$ observed from the point of view of the third reference state $\varrho$. In other words, the second requirement is essentially the small distance between near-equilibrium and equilibrium states observed by any third reference state. This formula of distance with the aid of a third reference state will be repeated in the following main results of this paper. If the reference state is selected as $\varrho=\rho_{ne}(\lambda_0,\eta)$, $\Delta S=S[\rho_{ne}(\lambda_0,\eta)||\rho_{e}(\lambda_0)]$, namely the direct distance of $\rho_{ne}(\lambda_0,\eta)$ and $\rho_{e}(\lambda_0)$. For high temperature where the second requirement of near-equilibrium state is satisfied, $S[\rho_{ne}(\lambda_0,\eta)||\rho_{e}(\lambda_0)]\approx\sum_m\rho'_{mm}(\lambda_0,0)\eta=0$. But for low temperature where the second requirement of near-equilibrium state is not satisfied, $S[\rho_{ne}(\lambda_0,\eta)||\rho_{e}(\lambda_0)]$ is larger than $1$, and even diverged.
In the derivation of $S[\rho_{ne}(\lambda_0,\eta)||\rho_{e}(\lambda_0)]\approx\sum_m\rho'_{mm}(\lambda_0,0)\eta$ above, we have used the property $\langle E_m(\lambda_0)|\psi'_m(\lambda_0,0)\rangle+\langle \psi'_m(\lambda_0,0)|E_m(\lambda_0)\rangle=0$.

\section{Quantum work distribution}\label{sec3}
Now, we consider the work distribution in the nonequilibrium process: At initial time $t=0$, the system is decoupled with the heat reservoirs and the weak priori driving field, then a protocol is performed on the system with the work parameter being changed from its initial value $\lambda_0$ to final value $\lambda_\tau$. The externally controlled evolution of the system is completely described by unitary operator
\begin{equation}\label{UU}
  U(\tau)\equiv\overleftarrow{T}\exp{\biggl[-i\int_{0}^{\tau}H(\lambda_t)dt\biggr]},
\end{equation}
where $\overleftarrow{T}$ is the time ordering operator, $H(\lambda_t)$ is the transient Hamiltonian of the system at time $t$, and $\tau$ is the duration of the protocol.

In order to determine the work done by such external control protocol, one needs to perform two energetic measurements at the beginning and the end of the external protocol. Traditionally, the first measurement is projective onto $\{|E_m(\lambda_0)\rangle\}$, i.e., the eigenstates of the initial Hamiltonian $H(\lambda_0)$. If the system is initially prepared in a state with quantum coherence, this first projective measurement will have a severe impact on the system dynamics and also on the work statistics through destroying quantum coherence. Thus, two-point measurement scheme is widely used on the initial incoherent state, e.g., thermal equilibrium state. But for the near-equilibrium state interested in this paper, there may be quantum coherence. We note that
\begin{equation}\label{}
  \langle\psi_m(\lambda_0,\eta)|H(\lambda_0)|\psi_m(\lambda_0,\eta)\rangle \approx E_m(\lambda_0)
\end{equation}
and
\begin{equation}\label{}
  \langle\psi_m(\lambda_0,\eta)|H(\lambda_0)|\psi_n(\lambda_0,\eta)\rangle \approx 0
\end{equation}
hold for any eigenstate of any near-equilibrium state, thus
\begin{equation}\label{}
  H(\lambda_0)|\psi_m(\lambda_0,\eta)\rangle\approx E_m(\lambda_0)|\psi_m(\lambda_0,\eta)\rangle.
\end{equation}
The first measurement can be chosen to project onto $\{|\psi_m(\lambda_0,\eta)\rangle\}$ instead of $\{|E_m(\lambda_0)\rangle\}$, and the corresponding energy is still $\{E_m(\lambda_0)\}$. This carefully selected first measurement has no effect on the initial state. After the first measurement, the system evolves under the unitary dynamics $U(\tau)$ generated by external protocol $\lambda_0\rightarrow\lambda_{\tau}$. Finally, the second projective measurement onto $\{|E_n(\lambda_\tau)\rangle\}$ is performed. One should prepare many copies of system with the same state $\rho_{ne}(\lambda_0,\eta)$, and then perform two-point measurement above for each copy. For each trial, one may obtain $E_m(\lambda_0)$ for the first measurement outcome followed by $E_n(\lambda_\tau)$ for the second measurement, and the corresponding probability is
\begin{equation}\label{}
  P(m,n)\approx\rho_{mm}(\lambda_0,\eta)P^{\tau}_{m\rightarrow n}
\end{equation}
with
\begin{equation}\label{}
  P^{\tau}_{m\rightarrow n}=\Big\vert\langle E_n(\lambda_\tau)|U(\tau)|\psi_m(\lambda_0,\eta)\rangle\Big\vert^2
\end{equation}
being the transfer probability from $|\psi_m(\lambda_0,\eta)\rangle$ to $|E_n(\lambda_\tau)\rangle$.
The work in the trajectory from $|\psi_m(\lambda_0,\eta)\rangle$ to $|E_n(\lambda_\tau)\rangle$ is defined as
\begin{equation}\label{}
  W_{ne}=E_n(\lambda_\tau)-E_m(\lambda_0)
\end{equation}
and whose probability distribution is
\begin{equation}\label{}
  P(W_{ne})\approx\sum_{mn}P(m,n)
\delta\Bigl(W_{ne}-\bigl(E_n(\lambda_\tau)-E_m(\lambda_0)\bigr)\Bigr).
\end{equation}
After Fourier transformation
\begin{equation}\label{}
\chi(\kappa)=\int P(W_{ne})e^{i\kappa W_{ne}}dW_{ne},
\end{equation}
the characteristic function of such quantum work distribution can be obtained as
\begin{equation}\label{chafunc}
  \chi(\kappa)\approx\mathrm{Tr}\Bigl[U^{\dag}(\tau)e^{i\kappa H(\lambda_{\tau})}U(\tau)\rho_{ne}(\lambda_0,\eta)
e^{-i\kappa H(\lambda_0)}\Bigr].
\end{equation}

\section{Quantum work fluctuation theorem}\label{sec4}
In this section, we derive the modified Jarzynski equality applicable to the initial near-equilibrium system.
Letting $\kappa=i\beta$ in Eq. (\ref{chafunc}), one can obtain
\begin{equation}\label{}
\langle e^{-\beta W_{ne}}\rangle
\approx\frac{Z(\lambda_\tau)}{Z(\lambda_0)}\Biggl[1+\mathrm{Tr}\biggl[U^{\dag}(\tau)\rho_{e}(\lambda_\tau)U(\tau)
\frac{\rho_{ne}(\lambda_0,\eta)-\rho_{e}(\lambda_0)}{\rho_{e}(\lambda_0)}\biggr]\Biggr].
\end{equation}
Since $\rho_{ne}(\lambda_0,\eta)-\rho_{e}(\lambda_0)\approx\rho'_{ne}(\lambda_0,0)\eta$ and $\rho_{ne}(\lambda_\tau,\eta)\approx\rho_{e}(\lambda_\tau)+\rho'_{ne}(\lambda_\tau,0)\eta$, the approximate formulas
\begin{equation}
  \mathrm{Tr}\biggl[U^{\dag}(\tau)\rho_{e}(\lambda_\tau)U(\tau)\frac{\rho_{ne}(\lambda_0,\eta)-\rho_{e}(\lambda_0)}{\rho_{e}(\lambda_0)}\biggr]
  \approx\mathrm{Tr}\biggl[U^{\dag}(\tau)\rho_{ne}(\lambda_\tau,\eta)U^{\dag}(\tau)
  \frac{\rho_{ne}(\lambda_0,\eta)-\rho_{e}(\lambda_0)}{\rho_{e}(\lambda_0)}\biggr]
\end{equation}
and
\begin{equation}\label{chi}
\langle e^{-\beta W_{ne}}\rangle\approx\frac{Z(\lambda_\tau)}{Z(\lambda_0)}\Biggl[1+
\mathrm{Tr}\biggl[U^{\dag}(\tau)\rho_{ne}(\lambda_\tau,\eta)U^{\dag}(\tau)
  \frac{\rho_{ne}(\lambda_0,\eta)-\rho_{e}(\lambda_0)}{\rho_{e}(\lambda_0)}\biggr]\Biggr]
\end{equation}
hold.
Because
\begin{equation}
  \bigg\vert\langle E_m(\lambda_0)|\frac{\rho_{ne}(\lambda_0,\eta)-\rho_{e}(\lambda_0)}{\rho_{e}(\lambda_0)}|E_m(\lambda_0)\rangle\bigg\vert
\approx\Bigg\vert\frac{\rho_{mm}(\lambda_0,\eta)-\frac{e^{-\beta E_m(\lambda_0)}}{Z(\lambda_0)}}{\frac{e^{-\beta E_m(\lambda_0)}}{Z(\lambda_0)}}\Bigg\vert<1
\end{equation}
holds for any $m$, $\frac{\rho_{ne}(\lambda_0,\eta)-\rho_{e}(\lambda_0)}{\rho_{e}(\lambda_0)}$ can be approximately expressed as
\begin{equation}
  \frac{\rho_{ne}(\lambda_0,\eta)-\rho_{e}(\lambda_0)}{\rho_{e}(\lambda_0)}\approx\ln\biggl[\mathbb{I}
  +\frac{\rho_{ne}(\lambda_0,\eta)-\rho_{e}(\lambda_0)}{\rho_{e}(\lambda_0)}\biggr]
  \approx\ln\rho_{ne}(\lambda_0,\eta)-\ln\rho_{e}(\lambda_0).
\end{equation}
Substituting it into Eq. (\ref{chi}), one can obtain
\begin{equation}\label{}
  \langle e^{-\beta W_{ne}}\rangle\approx\frac{Z(\lambda_\tau)}{Z(\lambda_0)}\biggl[1-\Delta S(\tau)\biggr],
\end{equation}
where
\begin{equation}
\begin{split}
\Delta S(\tau)&=\mathrm{Tr}\Bigl[\rho_{ne}(\lambda_\tau,\eta)U(\tau)\ln\rho_{e}(\lambda_0)U^\dag(\tau)\Bigr]
-\mathrm{Tr}\Bigl[\rho_{ne}(\lambda_\tau,\eta)U(\tau)\ln\rho_{ne}(\lambda_0,\eta)U^\dag(\tau)\Bigr] \\
&=S\Bigl[\rho_{ne}(\lambda_\tau,\eta)||U(\tau)\rho_{ne}(\lambda_0,\eta)U^\dag(\tau)\Bigr]
-S\Bigl[\rho_{ne}(\lambda_\tau,\eta)||U(\tau)\rho_{e}(\lambda_0)U^\dag(\tau)\Bigr]
\end{split}
\end{equation}
is the indirect distance between final states of $U(\tau)\rho_{ne}(\lambda_0,\eta)U^\dag(\tau)$ and $U(\tau)\rho_{e}(\lambda_0)U^\dag(\tau)$ observed from the point of view of reference state $\rho_{ne}(\lambda_\tau,\eta)$ [see Fig. \ref{figure1}(c)]. $\Delta S(\tau)$ characterises the difference between entropy productions caused by the external work protocol performing on near-equilibrium state $\rho_{ne}(\lambda_0,\eta)$ and equilibrium state $\rho_{e}(\lambda_0)$, respectively.
Because $\Delta S(\tau)\sim\eta$, $\ln[1-\Delta S(\tau)]\approx-\Delta S(\tau)$, and $Z(\lambda_\tau)/Z(\lambda_0)[1-\Delta S(\tau)]=\exp\{\ln[Z(\lambda_\tau)/Z(\lambda_0)]+\ln[1-\Delta S(\tau)]\}\approx\exp\{\ln[Z(\lambda_\tau)/Z(\lambda_0)]-\Delta S(\tau)\}=\exp\{-\beta\Delta F_{e}-\Delta S(\tau)\}$. Using this formula, the modified Jarzynski equality can be obtained as
\begin{equation}\label{MJE}
  \langle e^{-\beta W_{ne}}\rangle\approx e^{-\beta\Delta F_{e}-\Delta S(\tau)},
\end{equation}
which is the first main result of this paper.
This modified Jarzynski equality is a universal exact quantum work FT during a unitary process starting from near-equilibrium state, and includes all the information of quantum work statistics. Unlike the standard Jarzynski equality Eq. (\ref{JE}), which is independent of the process, the modified Jarzynski equality Eq. (\ref{MJE}) is process-dependent [shown in $\Delta S(\tau)$]. Such a process-dependent property is nontrivial that it gives a much tighter bound of work than the second law of thermodynamics, which will be discussed in detail below.

\section{The second law of thermodynamics}\label{sec5}
The modified Jarzynski equality can be expressed as the strict equality
\begin{equation}\label{}
 \langle e^{-\beta W_{ne}}\rangle=e^{-\beta\Delta F_{e}-\Delta S(\tau)-o(\eta^2)}.
\end{equation}
Using Jansen's inequality $\langle e^x\rangle\geq e^{\langle x\rangle}$, one can obtain
\begin{equation}\label{2thLaw1}
  \langle W_{ne}\rangle\geq\Delta F_e+T\Delta S(\tau)+o(\eta^2).
\end{equation}
Neglecting $o(\eta^2)$, Eq. (\ref{2thLaw1}) can be expressed as
\begin{equation}\label{2thLaw2}
  \langle W_{ne}\rangle\gtrsim\Delta F_e+T\Delta S(\tau),
\end{equation}
which is the second main result of this paper.
This result gives a fundamental constraint of work during a unitary process starting from near-equilibrium state.
It is worth emphasizing that since $\Delta S(\tau)$ is process-dependent, the work bound given by Eq. (\ref{2thLaw2}) is private for a given unitary process, which is different from the second law of equilibrium thermodynamics $\langle W_e\rangle\geq\Delta F_e$ [see Eq. (\ref{e2thlaw})] where the bound is public for all processes from the equilibrium state. What is the ultimate or public bound of the work for all processes from the near-equilibrium state? Using the property of the relative entropy $S(\rho||\tilde{\rho})\geq0$, one can find that
\begin{equation}\label{}
  \Delta S(\tau)\geq-\mathrm{Tr}\Bigl[\rho_{ne}(\lambda_\tau,\eta)\ln\rho_{ne}(\lambda_\tau,\eta)\Bigr]
  +\mathrm{Tr}\Bigl[\rho_{ne}(\lambda_\tau,\eta)U(\tau)\ln\rho_{e}(\lambda_0)U^\dag(\tau)\Bigr].
\end{equation}
In the near-equilibrium regime,
\begin{equation}\label{}
  \mathrm{Tr}\Bigl[\rho_{ne}(\lambda_\tau,\eta)\ln\rho_{ne}(\lambda_\tau,\eta)\Bigr]
\approx\mathrm{Tr}\Bigl[\rho_{ne}(\lambda_\tau,\eta)\ln\rho_{e}(\lambda_\tau)\Bigr],
\end{equation}
thus,
\begin{equation}\label{}
  \Delta S(\tau)\gtrsim-\mathrm{Tr}\Bigl[\rho_{ne}(\lambda_\tau,\eta)\ln\rho_{e}(\lambda_\tau)\Bigr]
  +\mathrm{Tr}\Bigl[\rho_{ne}(\lambda_\tau,\eta)U(\tau)\ln\rho_{e}(\lambda_0)U^\dag(\tau)\Bigr].
\end{equation}
The expression $-\mathrm{Tr}[\rho_{ne}(\lambda_\tau,\eta)\ln\rho_{e}(\lambda_\tau)]
+\mathrm{Tr}[\rho_{ne}(\lambda_\tau,\eta)U(\tau)\ln\rho_{e}(\lambda_0)U^\dag(\tau)]$ can be viewed as an integral
\begin{equation}\label{}
\begin{split}
-\mathrm{Tr}\Bigl[\rho_{ne}(\lambda_\tau,\eta)\ln\rho_{e}(\lambda_\tau)\Bigr]
  +\mathrm{Tr}\Bigl[\rho_{ne}(\lambda_\tau,\eta)U(\tau)\ln\rho_{e}(\lambda_0)U^\dag(\tau)\Bigr]=-\int \mathrm{Tr}\Bigl[\rho_{ne}(\lambda_\tau,\eta)d\ln\rho\Bigr]
\end{split}
\end{equation}
where the variation of state $\ln\rho$ is from $\ln U^\dag(\tau)\rho_{e}(\lambda_0)U(\tau)$ to $\ln \rho_{e}(\lambda_\tau)$.
I argue that one can always find an equivalent integral path $\Gamma$ of the variation of state $\rho(\lambda_t)$, depending on the work parameter $\lambda_t$, from $\rho_{ne}(\lambda_0,\eta)$ to $\rho_{ne}(\lambda_\tau,\eta)$, to make
\begin{equation}\label{}
\int_{\Gamma}\mathrm{Tr}\Bigl[\rho(\lambda_t)\frac{d\ln\rho_{e}(\lambda_t)}{d\lambda_t}\dot{\lambda}_t\Bigr]dt
=\mathrm{Tr}\Bigl[\rho_{ne}(\lambda_\tau,\eta)\ln\rho_{e}(\lambda_\tau)\Bigr]
  -\mathrm{Tr}\Bigl[U^\dag(\tau)\rho_{ne}(\lambda_\tau,\eta)U(\tau)\ln\rho_{e}(\lambda_0)\Bigr].
\end{equation}
Thus,
\begin{equation}
\begin{split}
S(\tau)&\gtrsim-\int_{\Gamma}\mathrm{Tr}\Bigl[\rho(\lambda_t)\frac{d\ln\rho_{e}(\lambda_t)}{d\lambda_t}
\dot{\lambda}_t\Bigr]dt \\
&=\beta\int_{\Gamma}\mathrm{Tr}\Bigl[\rho(\lambda_t)\frac{dH(\lambda_t)}{d\lambda_t}\dot{\lambda}_t\Bigr]dt
+\int_{\lambda_0}^{\lambda_\tau}\frac{d\ln Z(\lambda_t)}{d\lambda_t}d\lambda_t \\
&=\beta\int_{\Gamma}\mathrm{Tr}\Bigl[\rho(\lambda_t)\frac{dH(\lambda_t)}{d\lambda_t}\dot{\lambda}_t\Bigr]dt
+\ln\frac{Z(\lambda_\tau)}{Z(\lambda_0)} \\
&=\beta\int_{\Gamma}\mathrm{Tr}\Bigl[\rho(\lambda_t)\frac{dH(\lambda_t)}{d\lambda_t}\dot{\lambda}_t\Bigr]dt
-\beta\Delta F_e.
\end{split}
\end{equation}
Substituting it into Eq. (\ref{2thLaw1}) or Eq. (\ref{2thLaw2}), one can obtain
\begin{equation}\label{}
\langle W_{ne}\rangle\gtrsim\Delta F_e+T\Delta S(\tau)
  \gtrsim\int_{\Gamma}\mathrm{Tr}\Bigl[\rho(\lambda_t)\frac{dH(\lambda_t)}{d\lambda_t}
  \dot{\lambda}_t\Bigr]dt.
\end{equation}
The integral $\int_{\Gamma}\mathrm{Tr}[\rho(\lambda_t)\frac{dH(\lambda_t)}{d\lambda_t}\dot{\lambda}_t]dt$ can be understood as the average work along path $\Gamma$. It is worth emphasizing that this integral path $\Gamma$ is generally nonunitary. We assume this nonunitary process is generated by the external protocol performing on the system which is (1) coupled with two heat reservoirs at different temperatures [see Fig. \ref{figure1}(a)] or (2) is driven away from equilibrium by a weak field [see Fig. \ref{figure1}(b)], with the work parameter being changed from its initial value $\lambda_0$ to the final value $\lambda_\tau$. If the external protocol is infinitely slow that the system is always in the near-equilibrium state, we call this process as the quasistatic process, namely, near-equilibrium quasistatic process. In the equilibrium thermodynamics, it is well known that average work is minimized by the equilibrium quasistatic process, where the system is always in the equilibrium state at any time. Here, we argue that the average work during near-equilibrium quasistatic process is not greater than that of any finite time process from $\rho_{ne}(\lambda_0)$ to $\rho_{ne}(\lambda_\tau)$, i.e.,
\begin{equation}\label{}
\int_{\Gamma}\mathrm{Tr}\Bigl[\rho(\lambda_t)\frac{dH(\lambda_t)}{d\lambda_t}\dot{\lambda}_t\Bigr]dt
\geq\int_{\lambda_0}^{\lambda_\tau}
\mathrm{Tr}\Bigl[\rho_{ne}(\lambda_t,\eta)\frac{dH(\lambda_t)}{d\lambda_t}\Bigr]d\lambda_t.
\end{equation}
For this,
\begin{equation}\label{}
\langle W_{ne}\rangle\gtrsim\Delta F_e+T\Delta S(\tau)\geq\int_{\lambda_0}^{\lambda_\tau}
  \mathrm{Tr}\Bigl[\rho_{ne}(\lambda_t,\eta)\frac{dH(\lambda_t)}{d\lambda_t}\Bigr]d\lambda_t.
\end{equation}
In equilibrium thermodynamics, the average work during the equilibrium quasistatic process from $\rho_{e}(\lambda_0)$ to $\rho_{e}(\lambda_\tau)$ is defined as equilibrium free energy difference
\begin{equation}
\begin{split}
\Delta F_{e}=&\int_{\lambda_0}^{\lambda_\tau}\mathrm{Tr}\Bigl[\rho_e(\lambda_t)\frac{dH(\lambda_t)}{d\lambda_t}\Bigr]d\lambda_t \\
=&-T\int_{\lambda_0}^{\lambda_\tau}\frac{d\ln\mathrm{Tr}[e^{-\beta H(\lambda_t)}]}{d\lambda_t}d\lambda_t \\
=&-T\ln\frac{Z(\lambda_\tau)}{Z(\lambda_0)}.
\end{split}
\end{equation}
I assume the average work during the near-equilibrium quasistatic process from $\rho_{ne}(\lambda_0,\eta)$ to $\rho_{ne}(\lambda_\tau,\eta)$ is defined as near-equilibrium free energy difference, i.e.,
\begin{equation}\label{}
\int_{\lambda_0}^{\lambda_\tau}\mathrm{Tr}\Big[\rho_{ne}(\lambda_t,\eta)
\frac{dH(\lambda_t)}{d\lambda_t}\Big]d\lambda_t=\Delta F_{ne},
\end{equation}
thus, the third main result of this paper of
\begin{equation}\label{2thLaw}
  \langle W_{ne}\rangle\gtrsim\Delta F_e+T\Delta S(\tau)\geq\Delta F_{ne}
\end{equation}
can be obtained. In the equilibrium thermodynamics, it is well known that the work done on the system during a process from one equilibrium state to another must not be less than the accumulation of the equilibrium free energy of the system [see Eq. (\ref{e2thlaw})]. Here, Eq. (\ref{2thLaw}) demonstrates that this conclusion can be extended to non-equilibrium thermodynamics, i.e., the work done on the system during a process from one non-equilibrium state to another must still not be less than the accumulation free energy of the system, i.e.,
$\langle W_{ne}\rangle\geq\Delta F_{ne}$. Such an extension can be understood as the principle of maximum work or the second law of thermodynamics for near-equilibrium system. However, it should be noted that $\langle W_{ne}\rangle\gtrsim\Delta F_e+T\Delta S(\tau)$ offers a much tighter bound of work than the principle of maximum work.

\section{Two prototypical examples of near-equilibrium systems}\label{sec6}
In this section, two prototypical examples of near-equilibrium systems driven respectively by a temperature-gradient and an external field are taken into account to verify our main results.

\subsection{Temperature-gradient driving near-equilibrium system: Two-level system as an example}\label{sec61}
The simplest quantum system is a two-level system whose Hilbert space is spanned by just two states, an excited state $|e\rangle$ and a ground state $|g\rangle$. The Hamiltonian of the system is described as ($\hbar=1$)
\begin{equation}\label{}
  H_s=\frac{\omega_0}{2}\sigma_z,
\end{equation}
where $\sigma_z=|e\rangle\langle e|-|g\rangle\langle g|$ is the Pauli operator and $\omega_0$ is the transition frequency. For the convenience of discussion we let $\omega_0=1$. At time $t<0$, the two-level system is simultaneously coupled with two dissipative thermal baths with different temperatures. The dissipative dynamics of the two-level system is then generically determined by the following master equation \cite{Breuer2010}:
\begin{equation}\label{}
\begin{split}
  \frac{\partial\rho(t)}{\partial t}=-i[H_s,\rho(t)]
  &+\gamma_0\bigl(N_1(\omega_0)+N_2(\omega_0)+2\bigr)\Bigl[\sigma_-\rho(t)\sigma_+-\frac{1}{2}\bigl\{\sigma_+\sigma_-,\rho(t)\bigr\}\Bigr] \\
  &+\gamma_0\bigl(N_1(\omega_0)+N_2(\omega_0)\bigr)\Bigl[\sigma_+\rho(t)\sigma_--\frac{1}{2}\bigl\{\sigma_-\sigma_+,\rho(t)\bigr\}\Bigr],
\end{split}
\end{equation}
where $\gamma_0$ is the spontaneous dissipation rate and $N_j(\omega_0) = [\exp(\beta_j \omega_0)-1]^{-1}$ is the Planck distribution
with $\beta_j=1/T_j$ being the inverse temperature of the $j$th reservoir, and satisfies $N_j(-\omega_0) =-[1 + N_j(\omega_0)]$. The temperatures of two baths are assumed to be $T_1=T$ and $T_2=T+\Delta T$, respectively. After a sufficiently long time, the two-level system will stabilize at the near-equilibrium state for $\Delta T/T<1$, which is determined by solving equation $d\rho(t)/dt=0$. The near-equilibrium state can be expressed as the matrix form
\begin{equation}\label{rhone}
  \rho_{ne}(\omega_0,\Delta T/T)=\frac{1}{2\bigl(N_1(\omega_0)+N_2(\omega_0)+1\bigr)}\begin{pmatrix}N_1(\omega_0)+N_2(\omega_0) &0 \\ 0 & N_1(\omega_0)+N_2(\omega_0)+2\end{pmatrix}, 
\end{equation}
spanned by energy levels $|e\rangle$ and $|g\rangle$. 
If two thermal baths have the same temperature, i.e., $\Delta T=0$, the two-level system will stabilize at the equilibrium state, i.e.,
\begin{equation}\label{rhoe}
\lim_{\Delta T\rightarrow0}\rho_{ne}(\omega_0,\Delta T/T)=\frac{1}{2\cosh(\beta\omega_0/2)}
\begin{pmatrix}e^{-\frac{1}{2}\beta\omega_0} &0 \\ 0 & e^{\frac{1}{2}\beta\omega_0}\end{pmatrix}
=\frac{e^{-\beta H_s}}{\mathrm{Tr}[e^{-\beta H_s}]}=\rho_e(\omega_0).
\end{equation}

Once two-level system is stabilized at the near-equilibrium state (this time is labeled as $t=0$), it will be decoupled from two thermal baths, and driven by a time-dependent field. The Hamiltonian of the field-driven two-level system is
\begin{equation}\label{HST}
  H_s(t)=\frac{\omega_t}{2}\Bigl(\sigma_z\cos\frac{\pi t}{2\tau}+\sigma_x\sin\frac{\pi t}{2\tau}\Bigr),
\end{equation}
where $\hat{\sigma}_x=|e\rangle\langle g|+|g\rangle\langle e|$ is the Pauli operator, and $\omega_t=\omega_0(1-t/\tau)+\omega_{\tau}t/\tau$ is the linear ramp of the rf field frequency over time $\tau$, from $\omega_0$ to $\omega_{\tau}$, $t\in[0,\tau]$. The instantaneous Hamiltonian has a spectral decomposition $H_s(t)=\frac{\omega_t}{2}[|\psi_e(t)\rangle\langle\psi_e(t)|-|\psi_g(t)\rangle\langle\psi_g(t)|]$ with $|\psi_e(t)\rangle=\cos\frac{\pi t}{4\tau}|e\rangle+\sin\frac{\pi t}{4\tau}|g\rangle$ and $|\psi_g(t)\rangle=\sin\frac{\pi t}{4\tau}|e\rangle-\cos\frac{\pi t}{4\tau}|g\rangle$ being the instantaneous excited and ground states, respectively. The evolution of the system is governed  by unitary operator
\begin{equation}\label{}
  U_s(\tau)=\overleftarrow{T}\exp\Biggl[-i\int_0^{\tau}H_s(t)dt\Bigg],
\end{equation}
which can be obtained by numerically solving the ordinary differential equation $\dot{U}_s(t)=-iH_s(t)U_s(t)$ using 4th order Runge-Kutta. In order to verify our main results, we need to know the instantaneous near-equilibrium state with respect to $H_s(t)$. According to Eq. (\ref{rhone}), the instantaneous near-equilibrium state at time $t$ can be directly written, in the instantaneous energy basises $|\psi_e(t)\rangle$ and $|\psi_g(t)\rangle$, as
\begin{equation}
  \rho_{ne}(\omega_t,\Delta T/T)=\frac{1}{2\bigl(N_1(\omega_t)+N_2(\omega_t)+1\bigr)}\begin{pmatrix}N_1(\omega_t)+N_2(\omega_t) &0 \\ 0 & N_1(\omega_t)+N_2(\omega_t)+2\end{pmatrix}.
\end{equation}
The change of the near-equilibrium free energy is
\begin{equation}\label{}
\Delta F_{ne}=\int_{0}^{\tau}\mathrm{Tr}\bigg[\rho_{ne}(\omega_t)\frac{dH_s(t)}{dt}\bigg]dt
=-\frac{\omega_{\tau}-\omega_0}{2\tau}\int_{0}^{\tau}\frac{1}{N_1(\omega_t)+N_2(\omega_t)+1}dt.
\end{equation}

\begin{figure}
\begin{center}
\includegraphics[width=10cm]{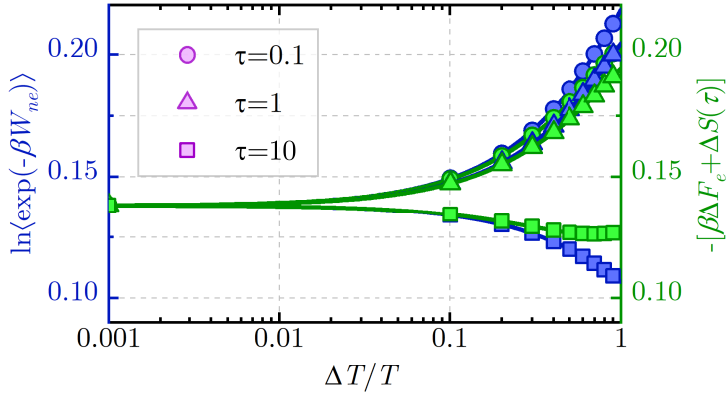}
\caption{ (Color online) The curves of $\ln\langle\exp(-\beta W_{ne})\rangle$ (blue) and $-[\beta\Delta F_e+\Delta S(\tau)]$ (olive) as the functions of $\Delta T/T$ are plotted for different driving duration $\tau=0.1$ (circle), $1$ (triangle) and $10$ (square). For all cases, $\omega_0=1$, $\omega_\tau=1.5$, and $T=1$.}
\label{figure2}
\end{center}
\end{figure}

In Fig. \ref{figure2}, we plot $\ln\langle\exp(-\beta W_{ne})\rangle$ and $-[\beta\Delta F_e+\Delta S(\tau)]$ as the functions of temperature difference for different driving duration. It can be seen that $\ln\langle\exp(-\beta W_{ne})\rangle$ and $-[\beta\Delta F_e+\Delta S(\tau)]$ are consistent with each other for $\Delta T/T<1$, and thus the modified Jarzynski equality $\langle\exp(-\beta W_{ne})\rangle\approx\exp[-\beta\Delta F_e-\Delta S(\tau)]$ holds. It is to be expected that $\ln\langle\exp(-\beta W_{ne})\rangle$ and $-[\beta\Delta F_e+\Delta S(\tau)]$ are inconsistent when temperature differences $\Delta T/T$ is increased that the system is far from equilibrium.

\begin{figure}
\begin{center}
\includegraphics[width=16cm]{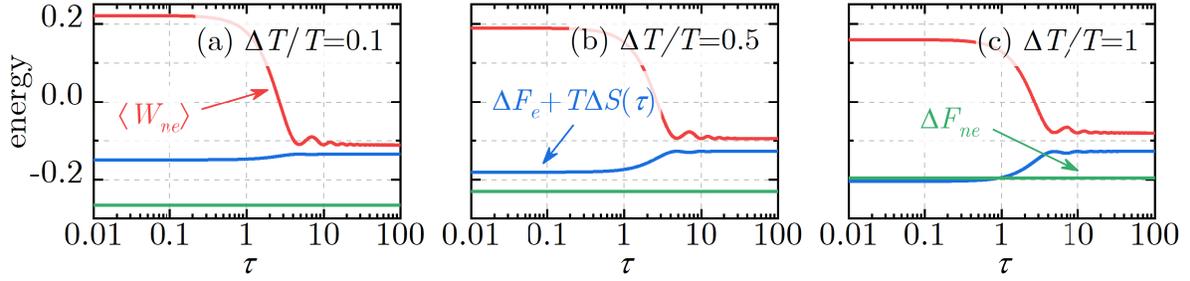}
\caption{ (Color online) The curves of $\langle W_{ne}\rangle$ (red) and $\beta\Delta F_e+\Delta S(\tau)$ (blue) and $\Delta F_{ne}$ (green) as the functions of driving duration $\tau$ for different temperature differences: (a) $\Delta T/T=0.1$, (b) $0.5$ and (c) $1$. For all cases, $\omega_0=1$, $\omega_\tau=1.5$, and $T=1$.}
\label{figure3}
\end{center}
\end{figure}

In order to verify the work constraint of Eq. (\ref{2thLaw}), we calculate $\langle W_{ne}\rangle$, $\Delta F_e+T\Delta S(\tau)$ and $\Delta F_{ne}$ as the function of driving duration $\tau$ for different temperature differences $\Delta T/T=0.1$, $0.5$ and $1$ (see Fig. \ref{figure3}). It can be seen that the work in the rapid driving process $\tau\ll1$ is much larger than the private bound $\Delta F_e+T\Delta S(\tau)$ and the public bound $\Delta F_{ne}$. In addition, slowing down the driving will make $\langle W_{ne}\rangle$ approach to the private bound of $\Delta F_e+T\Delta S(\tau)$. In other words, the work bound can be tightened by slowing down the driving. The work beyond free energy differences can be accounted for by average irreversible work $\langle W_{irr}\rangle=\langle W_{ne}\rangle-\Delta F_{ne}$, which will ultimately dissipate to the environment. In this sense, the rapid driving will waste more energy and reduce the efficiency of work done. The second law of thermodynamics for near-equilibrium system $\langle W_{ne}\rangle\gtrsim\Delta F_e+T\Delta S(\tau)\geq\Delta F_{ne}$ always holds for the small temperature differences [see Fig. \ref{figure3}(a) and (b)]. For the larger temperature differences that the system is far from equilibrium, although $\Delta F_{ne}$ can be greater than $\Delta F_e+T\Delta S(\tau)$, the inequality $\langle W_{ne}\rangle\geq\Delta F_{ne}$ always holds. This implies that for all the processes from one nonequilibrium steady state to another, the infinitely slow quasistatic process always do the smallest work, comparing with any finite-time process.

In the equilibrium thermodynamics, it is well known that the work done on the system during a process from one equilibrium state to another must not be less than the accumulation of the equilibrium free energy of the system, namely the principle of maximum work $\langle W_{e}\rangle\geq\Delta F_{e}$. The accumulation of equilibrium free energy is just the average work during the infinitely slow quasistatic process from one equilibrium state to another. The work beyond the equilibrium free energy difference, namely, the irreversible work, will ultimately dissipate to the environment. In this sense, compared with any finite-time process from one equilibrium state to another,   the infinitely slow quasistatic process is the most efficient because no extra energy is wasted. Here, the above conclusions in the equilibrium thermodynamics can be extended to the nonequilibrium systems that for all the processes from one nonequilibrium steady state to another, the infinitely slow quasistatic process is still the most efficient, comparing with any finite-time process. The principles of maximum work in both equilibrium thermodynamics and nonequilibrium thermodynamics have the same formulation that $\langle W\rangle\geq\Delta F$. If we drive the system from one steady state to another, whether they are equilibrium or not, the infinitely slow quasistatic process always do the smallest work.

Any realistic treatment of a quantum system cannot ignore the presence of the system's environment which is outside of our control. When the additional degrees of freedom of the environment are taken into account, the unitary evolution of a closed quantum system must be
modified to account for the system-environment interactions, which may cause e.g. decoherence and dissipation \cite{Meglio2023}. Now, we consider the driven-dissipative two-level system, i.e., while the two-level system is driven by the external field describing by Hamiltonian of Eq. (\ref{HST}), it also interacts with two thermal baths with different temperatures. We assume that the dynamics of the two-level system is determined by the following master equation
\begin{equation}\label{MSEQ}
\begin{split}
  \frac{d\rho(t)}{dt}=-i[H_s(t),\rho(t)]
  &+\gamma_0\bigl(N_1(\omega_0)+N_2(\omega_0)+2\bigr)\Bigl[\sigma_-\rho(t)\sigma_+-\frac{1}{2}\bigl\{\sigma_+\sigma_-,\rho(t)\bigr\}\Bigr] \\
  &+\gamma_0\bigl(N_1(\omega_0)+N_2(\omega_0)\bigr)\Bigl[\sigma_+\rho(t)\sigma_--\frac{1}{2}\bigl\{\sigma_-\sigma_+,\rho(t)\bigr\}\Bigr],
\end{split}
\end{equation}
which can be numerically solved by using 4th order Runge-Kutta. For this dynamics, it is challenging to define the trajectory work and heat, thus it is hard to verify the modified Jarzynski equality Eq. (\ref{MJE}) by this open quantum system. Extending the fluctuation theorems to the open quantum systems is our future work. However, average work during this process can be well defined as \cite{Ghosh2018}
\begin{equation}\label{}
  \langle W\rangle=\int_0^\tau\mathrm{Tr}[\rho(t)\dot{H}_s(t)]dt, 
\end{equation}
which can be used to verify the work constraint of Eq. (\ref{2thLaw}). Given the initial near-equilibrium state $\rho_{ne}(\omega_0,\Delta T/T)$ [see Eq. (\ref{rhone})] or equilibrium state $\rho_e(\omega_0)$ [see Eq. (\ref{rhoe})], the final state $\rho_{ne}(\tau)$ or $\rho_e(\tau)$ can be obtained by numerically solving Eq. (\ref{MSEQ}). In this case, $\Delta S(\tau)=S[\rho_{ne}(\omega_\tau)||\rho_{ne}(\tau)]-S[\rho_{ne}(\omega_\tau)||\rho_{e}(\tau)]$.

\begin{figure}
\begin{center}
\includegraphics[width=16cm]{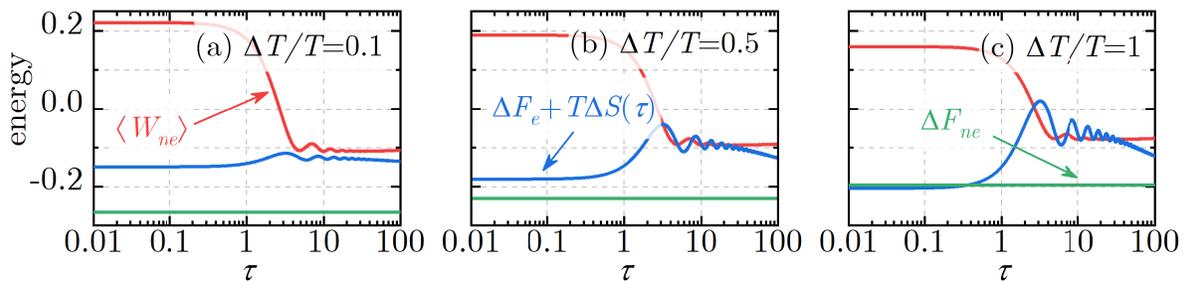}
\caption{ (Color online) The curves of $\langle W_{ne}\rangle$ (red) and $\beta\Delta F_e+\Delta S(\tau)$ (blue) and $\Delta F_{ne}$ (green) as the functions of driving duration $\tau$ for different temperature differences: (a) $\Delta T/T=0.1$, (b) $0.5$ and (c) $1$. For all cases, $\omega_0=1$, $\omega_\tau=1.5$, $\gamma_0=0.01$ and $T=1$.}
\label{figure4}
\end{center}
\end{figure}

In Fig. \ref{figure4}, we plot $\langle W_{ne}\rangle$, $\Delta F_e+T\Delta S(\tau)$ and $\Delta F_{ne}$ as the function of driving duration $\tau$ for different temperature differences $\Delta T/T=0.1$, $0.5$ and $1$. Similar to driven isolated two-level system (see Fig. \ref{figure3}), $\langle W_{ne}\rangle$ in the rapid driving process $\tau\ll1$ is much larger than the private bound of work $\Delta F_e+T\Delta S(\tau)$ and the public bound $\Delta F_{ne}$. Slowing down the driving, the work $\langle W_{ne}\rangle$ will approach to the private bound of $\Delta F_e+T\Delta S(\tau)$. The second law of thermodynamics for near-equilibrium system $\langle W_{ne}\rangle\gtrsim\Delta F_e+T\Delta S(\tau)\geq\Delta F_{ne}$ always holds for the small temperature differences. Although $\Delta F_{ne}$ can be greater than $\Delta F_e+T\Delta S(\tau)$ for large temperature differences, the inequality $\langle W_{ne}\rangle\geq\Delta F_{ne}$ always holds.

\subsection{External field driving near-equilibrium: Quantum Ising model as an example}\label{sec62}
Recently, to study the dynamics of quantum many-body systems in this nonequilibrium thermodynamical formulation has aroused widespread interest. There has been quite a remarkable amount of activity uncovering the features of work statistics in a range of physical models including spin chains \cite{Silva2008,Smacchia2013,Fusco2014,Mascarenhas2014,Dorner2012,Arrais2019}, Fermionic systems \cite{Heyl2012,Schiro2014,Vicari2019,Zawadzki2020}, Bosonic systems and Luttinger liquids \cite{Lena2016,Villa2018,Dora2013,Basci2013} and periodically driven quantum systems \cite{Dutta2015,Russomanno2015,Bunin2011,Russomanno2012}. Work statistics have also proved to be useful in the analysis of dynamical quantum criticality \cite{Heyl2013,Heyl2018,Quan2017,Fei2020,Zhang2022,Mzaouali2021} and more recently to shed light on the phenomenon of information scrambling \cite{Campisi2017,Chenu2018,Tsuji2018,Chenu2019}. Let me now specialize our discussions to the one-dimensional quantum Ising chain. The Hamiltonian considered is
\begin{equation}\label{HIsing}
  H\bigl(\lambda_t\bigr)=-\frac{J}{2}\sum_{j=1}^{L}\bigl[\sigma_j^z\sigma_{j+1}^z+\lambda_t\sigma_j^x\bigr],
\end{equation}
where $\sigma_j^{x,z}$ are the spin operators at lattice site $j$, $\lambda_t$ is the time dependent transverse field, $L$ is the length of the Ising chain and $J$ is longitudinal coupling. In this work, we set $J=1$ as the overall energy scale. The one-dimensional quantum Ising model is the prototypical, exactly solvable example of a quantum phase transition \cite{Sachdev2011}, with a quantum critical point at $\lambda_c=1$ separating a quantum paramagnetic phase at $\lambda>\lambda_c$ from a ferromagnetic one at $\lambda<\lambda_c$.

After Jordan-Wigner transformation and Fourier transforming, Hamiltonian Eq. (\ref{HIsing}) becomes a sum of two-level systems \cite{Russomanno2012}:
\begin{equation}\label{}
\hat{H}(\lambda_t)=\sum_k\hat{H}_k(\lambda_t).
\end{equation}
Each $\hat{H}_k(\lambda_t)$ acts on a two-dimensional Hilbert space generated by $\{\hat{c}^\dag_k\hat{c}^\dag_{-k}|0\rangle,~|0\rangle\}$, where $|0\rangle$ is the vacuum of the Jordan-Wigner fermions $\hat{c}_k$, and can be represented in that basis by a $2\times2$ matrix
\begin{equation}\label{}
 \hat{H}_k(\lambda_t)=\epsilon_k(\lambda_t)\sigma^z+\Delta_k\sigma^y,
\end{equation}
where $\epsilon_k(\lambda_t)=\lambda_t-\cos k$ and $\Delta_k=\sin k$ and $k=(2n-1)\pi/L$ with $n=1\cdots L/2$, corresponding to antiperiodic boundary conditions for $L$ is even. The instantaneous eigenvalues are
\begin{equation}\label{}
\varepsilon^{\pm}_{k}(\lambda_t)=\pm\varepsilon_{k}(\lambda_t)=\pm\sqrt{\epsilon^2_k(\lambda_t)+\Delta^2_k},
\end{equation}
and the corresponding eigenvectors are
\begin{equation}\label{}
|\psi^+_{k}(\lambda_t)\rangle=
\bigl[\cos\theta_k+i\sin\theta_k\hat{c}^\dag_k\hat{c}^\dag_{-k}\bigr]|0\rangle
\end{equation}
and
\begin{equation}\label{}
|\psi^-_{k}(\lambda_t)\rangle=\bigl[i\sin\theta_k+\cos\theta_k\hat{c}^\dag_k\hat{c}^\dag_{-k}\bigr]|0\rangle,
\end{equation}
respectively, where $\theta_k=\arctan[\frac{\Delta_k}{\epsilon_k(\lambda_t)-\varepsilon_{k}(\lambda_t)}]$.

In order to understand quantum work FT of a process starting from the near-equilibrium state, we consider that at time $t<0$ the system is driven by a time independent field $\lambda'_0=\lambda_0+\delta_\lambda$ and coupled with one heat reservoir at temperature $T$. In this case, $\eta=\delta_\lambda/J$. After thermalization, the state of the system can be described by $\rho_{e}(\lambda'_0)\equiv\exp\{-\beta\hat{H}(\lambda'_0)\}/Z(\lambda'_0)=\bigotimes_{k}\exp\{-\beta\hat{H}_k(\lambda'_0)\}/Z_k(\lambda'_0)$ with $Z_k(\lambda'_0)=2\cosh(\sqrt{\epsilon^2_k(\lambda'_0)+\Delta^2_k})$ being the partition function of the mode $k$. At time $t=0$, the system is decoupled with the heat reservoir and the time independent field is suddenly changed to the linearly time-dependent field
\begin{equation}\label{}
  \lambda_t=\lambda_0+\frac{\lambda_\tau-\lambda_0}{\tau}t.
\end{equation}
Then, we will investigate the statistical properties of the work performed by this linearly time-dependent field. The initial state of the system is near-equilibrium for the initial Hamiltonian $H(\lambda_0)$, i.e., $\rho_{ne}(\lambda_0,\eta)=\rho_{e}(\lambda'_0), $ because $\delta_\lambda$ is weak.

\begin{figure}
\begin{center}
\includegraphics[width=10cm]{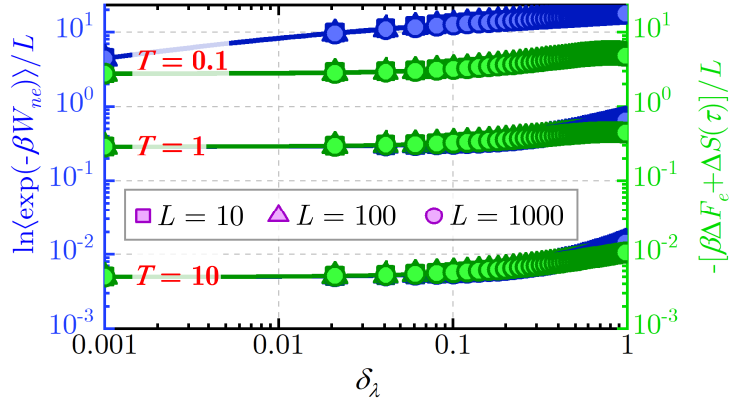}
\caption{ (Color online) The curves of $\ln\langle\exp(-\beta W_{ne})\rangle/L$ (blue) and $-[\beta\Delta F_e+\Delta S(\tau)]/L$ (olive) as the functions of $\delta_{\lambda}$ for different length of Ising chain $L=10$ (square), $L=100$ (triangle) and $L=1000$ (circle) are plotted at $T=0.1$ (low temperature), $T=1$ (medium temperature) and $T=10$ (high temperature). Given a temperature, whether $\ln\langle\exp(-\beta W_{ne})\rangle/L$ or $-[\beta\Delta F_e+\Delta S(\tau)]/L$, their curves for different length of Ising chain $L=10$ (square), $L=100$ (triangle) and $L=1000$ (circle) coincide with each other. For all cases, $\lambda_0=0$, $\lambda_\tau=\lambda_0+1$, and $\tau=1$.}
\label{figure5}
\end{center}
\end{figure}

The time evolution of the system state caused by linear time dependent field has a BCS-like form $|\Psi(t)\rangle=\bigotimes_{k}|\psi_k(t)\rangle=\bigotimes_{k}[v_k(t)+u_k(t)\hat{c}^\dag_k\hat{c}^\dag_{-k}]|0\rangle$. The state coefficients $\{v_k(t),~u_k(t)\}$ are given by solution of the Bogoliubov-de Gennes equations ($\hbar=1$):
\begin{equation}\label{Schrodinger equation}
\begin{pmatrix}\dot{u}_k(t) \\ \dot{v}_k(t) \end{pmatrix}
=-i\hat{H}_k\bigl(\lambda_t\bigr)\begin{pmatrix}u_k(t) \\ v_k(t) \end{pmatrix}.
\end{equation}
This equation can be numerically solved by using 4th order Runge-Kutta. It can be verified that if $(u_k(t),v_k(t))^t$ solves the Bogoliubov-de Gennes equations with initial condition $(1,0)^t$, also $(-v^\ast_k(t),u^\ast_k(t))^t$ is the solution but with initial condition $(0,1)^t$. Therefore the time evolution operator $\hat{U}_k(t,0)=\overleftarrow{T}\exp[-i\int_0^tdt'\hat{H}_k(\lambda_{t'})]$ can be written as \cite{Russomanno2012}
\begin{equation}\label{}
  \hat{U}_k(t,0)=\begin{pmatrix}u_k(t) &-v^\ast_k(t) \\ v_k(t) & u^\ast_k(t)\end{pmatrix}.
\end{equation}
At this point we have translated the many-body problem into the solution of the time-dependent Schr\"{o}dinger equation for an effective two-level system.

\begin{figure}
\begin{center}
\includegraphics[width=10cm]{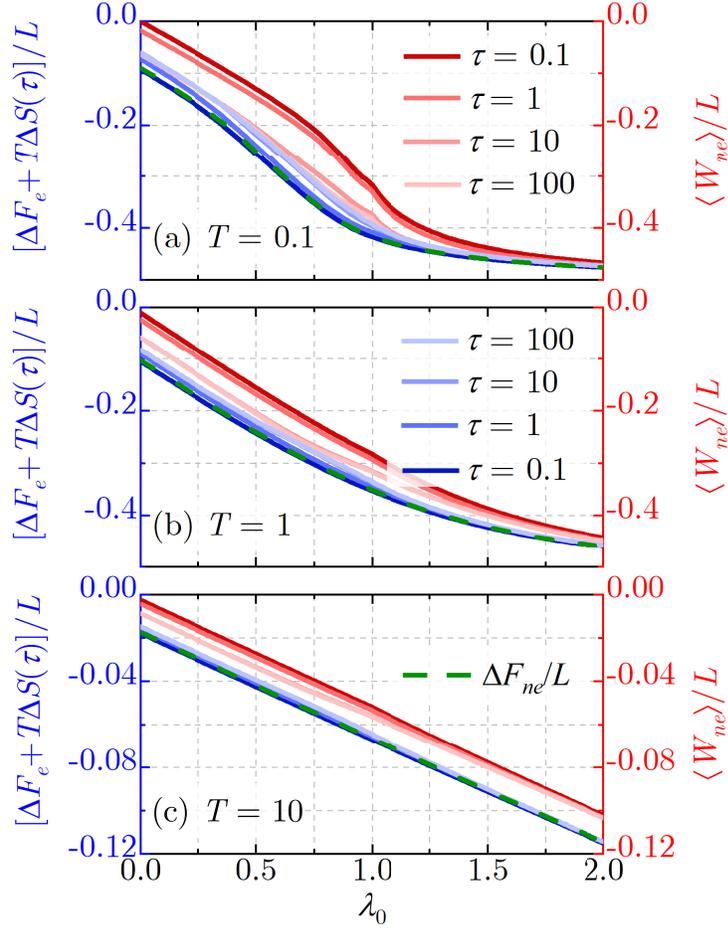}
\caption{ (Color online) The curves of $\langle W_{ne}\rangle/L$ (red) and $[\Delta F_e+T\Delta S(\tau)]/L$ (blue) as the functions of $\lambda_0$ for different duration of the protocol $\tau=0.1$, $1$, $10$ and $100$ are plotted at (a) $T=0.1$ (low temperature), (b) $T=1$ (medium temperature) and (c) $T=10$ (high temperature). The reference curves (olive dashed) are $\Delta F_{ne}$. For all cases $\delta_{\lambda}=0.1J$, $\lambda_\tau=\lambda_0+0.5$, $L=100$.}
\label{figure6}
\end{center}
\end{figure}

In Fig. \ref{figure5}, $\ln\langle\exp(-\beta W_{ne})\rangle/L$ and $-[\beta\Delta F_e+\Delta S(\tau)]/L$ for different length of Ising chain are plotted at different temperatures. Interestingly, one can find that given a temperature, they both converge to their own certain values as the length of Ising chain increases, because the energy density is an intensive quantity. After comparison, it can be found that when temperature is not low, i.e., $T\geq1$, $\ln\langle\exp(-\beta W_{ne})\rangle/L$ and $-[\beta\Delta F_e+\Delta S(\tau)]/L$ are consistent with each other for $\delta_{\lambda}<1$, and thus the modified Jarzynski equality $\langle\exp(-\beta W_{ne})\rangle\approx\exp[-\beta\Delta F_e-\Delta S(\tau)]$ holds. Just as the analyses in Sec. \ref{sec2}, at low temperature $T<1$ (e.g., $T=0.1$), the system is so sensitive to external perturbation that it is hard to satisfy the condition of near-equilibrium state, thereby $\ln\langle\exp(-\beta W_{ne})\rangle/L$ and $-[\beta\Delta F_e+\Delta S(\tau)]/L$ are inconsistent.

\begin{figure}
\begin{center}
\includegraphics[width=9cm]{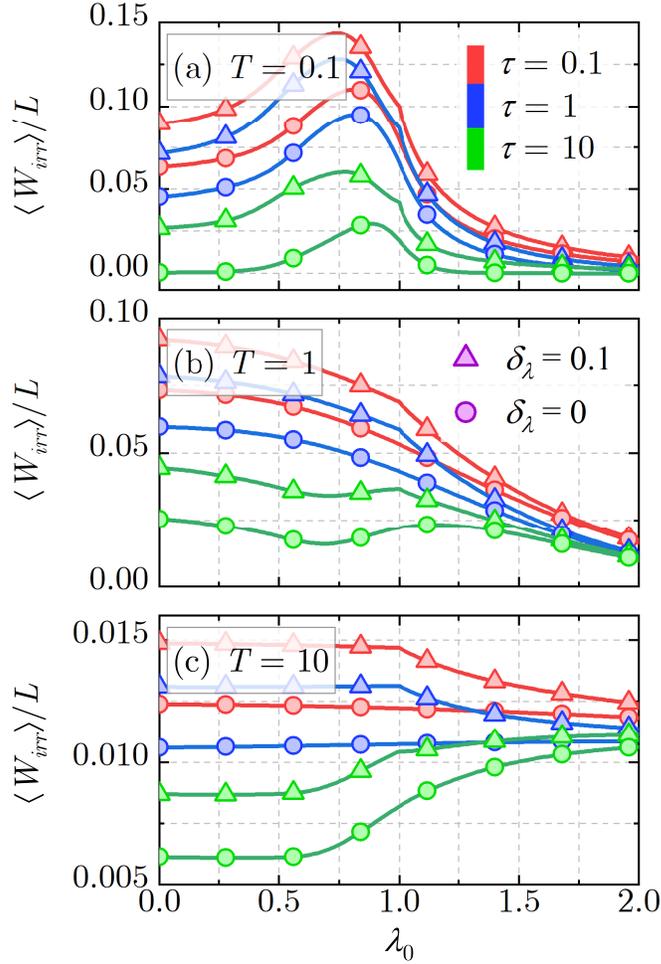}
\caption{ (Color online) (a) The curves of near-equilibrium (triangle, $\delta_\lambda=0.1$) and equilibrium (circle, $\delta_\lambda=0$) irreversible work densities $\langle W_{irr}\rangle/L$ as the functions of $\lambda_0$ for different duration of the protocol $\tau=0.1$ (red), $1$ (blue) and $10$ (green) are plotted at (a) $T=0.1$ (low temperature), (b) $T=1$ (medium temperature) and (c) $T=10$ (high temperature). For all cases, $\lambda_\tau=\lambda_0+0.5$ and $L=1000$. }
\label{figure7}
\end{center}
\end{figure}

In order to verify the work constraint of Eq. (\ref{2thLaw}), we calculate $\langle W_{ne}\rangle/L$ and $[\Delta F_e+T\Delta S(\tau)]/L$ for different duration of the protocol at different temperatures, and compare them with the density of the near-equilibrium free energy $\Delta F_{ne}/L=-T\ln[Z(\lambda'_\tau)/Z(\lambda'_0)]/L$ (see Fig. \ref{figure6}). It can be seen that the second law of thermodynamics for near-equilibrium system $\langle W_{ne}\rangle\gtrsim\Delta F_e+T\Delta S(\tau)\geq\Delta F_{ne}$ always holds, even at low temperature $T=0.1$ where the system is far away from equilibrium. For the rapid driving $\tau=0.1$, the private bound of work $\Delta F_e+T\Delta S(\tau)$ is consistent with the public bound $\Delta F_{ne}$ (i.e., free energy differences), and the work in this rapid driving process is lager than them. $\Delta F_e+T\Delta S(\tau)$ will be greater than $F_{ne}$ if driving is slowed, but $\langle W_{ne}\rangle$ will approach to $\Delta F_e+T\Delta S(\tau)$, especially at low temperature. The rapid driving will generate more irreversible work $\langle W_{irr}\rangle=\langle W_{ne}\rangle-\Delta F_{ne}$ and reduce the efficiency of work done, which is similar to the result of two-level system. The similarity implies the universality of this result. As the length of Ising chain approaches to the thermodynamic limit ($L\rightarrow\infty$), the density of the irreversible work $\langle W_{irr}\rangle/L$ converges to a certain value which is independent of $L$.

In general, quantum phase transitions are held to occur only at zero temperature because thermal fluctuation can destroy quantum phase transition or the "sea" of thermal fluctuations will drown the information of quantum phase transition. In the zero temperature limit where the system is in the ground state, work is performed to drive the system across the critical region and, due to the vanishing energy gap, it becomes increasingly difficult to do so without exciting system, thereby sharpening the irreversible work \cite{Dorner2012}. This leads to the production of irreversible entropy and the emergence of intrinsic irreversibility in the critical region. But if the driving is not very weak (e.g., $\Delta\lambda=\lambda_\tau-\lambda_0=0.5$, which is much larger than $\Delta\lambda=0.01$ in Ref. \cite{Dorner2012}), the sharpening behavior of the irreversible work will disappear [see the circle marked curves in Fig. \ref{figure7}(a)]. If the system is pushed out of equilibrium by a priori perturbation before driving, a kink can be observed at the critical point $\lambda_c=1$ [see the triangle marked curves in Fig. \ref{figure7}(a)]. This singularity is so stable that the thermal fluctuations can not destroy it [see the triangle marked curves in Fig. \ref{figure7}(b) and (c)]! This may imply that quantum phase transition can be observed and investigated at high temperature, the only need is to employ a perturbation to push the system out of equilibrium, even a little. After this perturbation, the information of quantum phase transition can be recaptured from "sea" of thermal fluctuations through the system's responses.

\section{Conclusions}

To summarize, this paper has considered a quantum system is initially driven from equilibrium to a near-equilibrium state by a priori small perturbation, e.g., small temperature differences or weak external field, then a general post work protocol is performed on it. Using perturbation theory, the work FT and the corresponding second law of thermodynamics for the initial near-equilibrium state has been derived. Our results gave a much tighter bound of work for a given process than the second law of thermodynamics. Finally, we considered a two-level system and a transverse field quantum Ising model to examine the main results. Except for the validity of our results, a very rare critical phenomenon at really high temperature has been found. This may pave a new way to investigate quantum phase transition, i.e., the information of quantum phase transition drowned out by thermal noise can be recaptured through the system's response after a perturbation.

Recently, a general Jarsynski equality for arbitrary initial state was proposed within a resource theoretic framework in Ref. \cite{Alhambra2016_1}. A natural question then arises: Whether the main results of the modified Jarzynski equality [Eq. (\ref{MJE})] and the second law of thermodynamics for near-equilibrium system [Eq. (\ref{2thLaw})] can be derived from the resource theory results in Ref. \cite{Alhambra2016_1}? That is not the case because the paradigm of resource theory is completely different from this paper. It does not apply to the research framework used in this paper, let alone get the results derived from it.

In resource theory, the thermodynamical transition is simulated by thermal operation where the interactions between the system, heat bath, work storage device and the switch system are carefully designed, and some other additional maps that depend on the initial state are applied. These interactions and additional maps will inevitably affect the system, which can be understood as some effective measurements. Thus resource theory can not cleanly simulate the system dynamics, e.g., time-dependent driving and sudden quench, without any other influences, and the results derived from it will include these additional influences above. If one wants to eliminate these influences, the initial state of the system should be restricted, for example, to the thermal equilibrium state, rather than arbitrary state.

Unlike resource theory, the system dynamics where the work parameter is changing from its initial value $\lambda_0$ to the final value $\lambda_\tau$ or the system Hamiltonian is changing from $H(\lambda_0)$ to $H(\lambda_\tau)$ was completely described by a unitary operator Eq. (\ref{UU}) in this paper. Based on this framework alone and without considering other contributions, the fluctuation work was defined, and a modified Jarzynski equality and the corresponding second law applicable to the initial near-equilibrium state were derived. The simplicity of the derived approach should moreover enable future extensions to open system dynamics.

\section*{Acknowledgment}
I am grateful to Profs. Jian-Hui Wang, Mang Feng, Fei Liu and Jian Zou for helpful comments and discussions. This work was supported by the National Natural Science Foundation of China (Grants No. 11705099) and the Talent Introduction Project of Dezhou University of China (Grant No. 30101437).

\end{document}